\documentclass[a4paper,12]{article}
\usepackage{amsmath,amsfonts,amssymb,graphicx,cite}
\usepackage[paper=letterpaper,margin=1.2in]{geometry}

\begin{document}

{\footnotesize
\rightline{VPI-IPNAS-09-03}
}

\vskip 1.5 cm
\renewcommand{\thefootnote}{\fnsymbol{footnote}}
\centerline{\Large \bf Eigenvalue Density, Li's Positivity, and the Critical Strip}
\vskip 0.75 cm

\centerline{{\bf
Yang-Hui He${}^{1}$\footnote{\tt hey@maths.ox.ac.uk},
Vishnu Jejjala${}^{2}$\footnote{\tt vishnu@ihes.fr},
and Djordje Minic${}^{3}$\footnote{\tt dminic@vt.edu}
}}
\vskip .5cm
\centerline{${}^1$\it Collegium Mertonense in Universitate Oxoniensi, OX1 4JD;}
\centerline{\it Rudolf Peierls Centre for Theoretical Physics,}
\centerline{\it Oxford University, 1 Keble Road, OX1 3NP; \&}
\centerline{\it Mathematics Institute, Oxford University,}
\centerline{\it 24-29 St.\ Giles', Oxford, OX1 3LB, U.K.}
\vskip .5cm
\centerline{${}^2$\it Institut des Hautes \'Etudes Scientifiques,}
\centerline{\it 35, Route de Chartres, 91440 Bures-sur-Yvette, France}
\vskip .5cm
\centerline{${}^3$\it Institute for Particle, Nuclear, and Astronomical Sciences,}
\centerline{\it Department of Physics, Virginia Tech,}
\centerline{\it Blacksburg, VA 24061, U.S.A.}
\vskip .5cm

\newcommand{\im}{\mathop{{\rm Im}}}
\newcommand{\re}{\mathop{{\rm Re}}}
\newcommand{\tr}{\mathop{{\rm Tr}}}
\newcommand{\cO}{{\cal O}}
\newcommand{\cS}{{\cal S}}
\newcommand{\cL}{{\cal L}}
\newcommand{\cT}{{\cal T}}
\newcommand{\cH}{{\cal H}}
\newcommand{\nn}{{\nonumber}}
\newcommand{\IR}{\mathbb{R}}
\newcommand{\IC}{\mathbb{C}}
\newcommand{\IZ}{\mathbb{Z}}
\newcommand{\id}{\mathbb{I}}
\newcommand{\comment}[1]{}

\newcommand{\be}{\begin{equation}}
\newcommand{\ee}{\end{equation}}
\newcommand{\bea}{\begin{eqnarray}}
\newcommand{\eea}{\end{eqnarray}}
\newcommand{\ba}{\begin{array}}
\newcommand{\ea}{\end{array}}
\newcommand{\ben}{\begin{enumerate}}
\newcommand{\een}{\end{enumerate}}
\newcommand{\bi}{\begin{itemize}}
\newcommand{\ei}{\end{itemize}}
\newcommand{\bc}{\begin{center}}
\newcommand{\ec}{\end{center}}
\newcommand{\bt}{\begin{table}}
\newcommand{\et}{\end{table}}
\newcommand{\btab}{\begin{tabular}}
\newcommand{\etab}{\end{tabular}}
\newcommand{\bs}{\begin{slide}}
\newcommand{\es}{\end{slide}}
\newcommand{\eref}[1]{(\ref{#1})}
\newcommand{\todo}[1]{\noindent{\em #1}\marginpar{$\Longleftarrow$}}
\newcommand{\dd}{{\rm d}}

\newtheorem{theorem}{\bf THEOREM}
\def\thetheorem{\thesection.\arabic{theorem}}
\newtheorem{conjecture}{\bf CONJECTURE}
\def\thetheorem{\thesection.\arabic{conjecture}}
\newtheorem{proposition}{\bf PROPOSITION}
\def\thetheorem{\thesection.\arabic{proposition}}

\def\theequation{\thesection.\arabic{equation}}
\newcommand{\setall}{\setcounter{equation}{0}\setcounter{theorem}{0}\setcounter{table}{0}\setcounter{footnote}{0}}
\newcommand{\setequation}{\setcounter{equation}{0}}

\begin{abstract}
We rewrite the zero-counting formula within the critical strip of the Riemann zeta function as a cumulative density distribution;
this subsequently allows us to formally derive an integral expression for the Li coefficients associated with the Riemann $\xi$-function which, in particular, indicate that their positivity criterion is obeyed, whereby entailing the criticality of the non-trivial zeros. We conjecture the validity of this and related expressions without the need for the Riemann Hypothesis and
also offer a physical interpretation of the result and discuss the Hilbert--P\'olya approach.
\end{abstract}

\newpage

\tableofcontents

\section{Introduction}\setall
The zeros of the Riemann zeta function have been extensively studied in mathematics.
Trivial zeros of $\zeta(s)$ occur at the negative even integers.
In 1859, Riemann~\cite{riemann} conjectured that all non-trivial zeros have a real part $\frac12$.
See~\cite{bombieri, conrey} for nice reviews.

The idea to apply physics to the Riemann zeros is an old one.
Hilbert and P\'olya~\cite{hp} independently proposed that non-trivial zeros of the Riemann zeta function could correspond to eigenvalues of a self-adjoint (Hermitian) operator.
Subsequent research is suggestive of this {\em spectral interpretation} of the zeros of the zeta function.
The Selberg trace formula~\cite{selberg} encapsulates the duality between the lengths of geodesics on a Riemann surface and the spectrum of the Laplacian on the surface.
For the Laplacian acting on the space of $PSL(2,\IZ)$ invariant real analytic functions on the upper half plane the eigenvalue distribution is related to the Selberg zeta function ${\cal Z}(s)$ in the same way that Weil's explicit formula~\cite{weil} from analytic number theory is related to the Riemann zeta function.

Subsequently, Montgomery~\cite{monty} observed that the pair correlation function of zeros of the zeta function along the critical line follows the distribution $X(u) \sim 1 - \left( \frac{\sin\pi u}{\pi u} \right)^2$.
This gives the density of spacings of not necessarily consecutive zeros.
Dyson realized that the density $X(u)$ specifies the pair correlation function of the eigenvalues of a large random Hermitian matrix with a Gaussian measure.
The statistics of the zeros of the zeta function therefore follow the {\em Gaussian unitary ensemble}, a fact verified spectacularly through the numerical studies of Odlyzko~\cite{odlyzko}.

Over the years the spectral interpretation has been explored by various researchers in an effort to bring quantum physics
and statistical physics to bear upon this difficult mathematical problem.
A large list of references is included in~\cite{www}.
More recently the connection between physics and the Riemann zeros has been explored in the works of Berry and Keating~\cite{berry} and its subsequent elaborations by Connes~\cite{connes} and by Sierra and Townsend~\cite{sierra} regarding the physical interpretation of the formula for the number of non-trivial zeros in a given interval.

The aforementioned formula can be written explicitly in terms of the Riemann $\xi$-function, which was also used by Li~\cite{li, bl} to establish an equivalent formulation of the Riemann Hypothesis.
The present article attempts to establish a connection between the physics of the exact formula for the number of non-trivial zeros $N(T)$ up to a given height in the critical strip and Li's positivity criterion.
Our arguments rely on the property of $N(T)$ as a cumulative density function.
The intuition and our discussion find their wellspring in theoretical physics.

The note is organized as follows.
First, we review Li's Criterion, with an emphasis on requisite properties of the Riemann $\xi$-function.
In the central section of the paper, we show how using $N(T)$ and its derivative as eigenvalue distributions in the critical strip would imply the positivity of Li's coefficients.
Finally, we interpret the result from a physicist's perspective.

\subsection*{Nomenclature}
Throughout this paper, we will adhere to the following notation.
\vskip .5cm
\begin{tabular}{ll}
$\xi(z)$ & The Riemann $\xi$-function,
$\xi(s) := \frac12 s(s-1) \pi^{-\frac{s}{2}} \Gamma(s/2) \zeta(s)$. \\
$\cS$ & The critical strip,
$\cS := \{z \ | \ 0 < \re(z) < 1 \}$. \\
$\cS_+$ & The upper critical strip,
$\cS_+ := \{z \ | \ 0 < \re(z) < 1,\ \im(z) > 0 \}$. \\
$\cL$ & The critical line,
$\cL := \{ z \ | \ \re(z) = \frac12 \}$. \\
$\cL_+$ & The upper critical line,
$\cL := \{ z \ | \ \re(z) = \frac12,\ \im(z) > 0\}$. \\
$N(T)$ & The number of zeros of $\zeta(z)$ or $\xi(z)$ inside $\cS_+$ up to height $\im(z) < T$.\\
$\rho_{j,k} = \sigma_{j,k} + i \mu_j$ & A zero of $\zeta(z)$ or $\xi(z)$ in $\cS_+$ indexed by $j,k\in \IZ_{>0}$; $\mu_{j+1} > \mu_j$, $\sigma_{j,k+1} \ge \sigma_{j,k}$.
\end{tabular}

\section{Li's Positivity Criterion}\setall
Let us, perhaps more for the sake of notation, first remind the reader of the statement of the Riemann Hypothesis~\cite{riemann, e, t, ivic}:
\begin{quote}{\it
The analytic continuation, from $z \in \IR_{>1}$ to the whole complex plane $z \in \IC$ of
$\zeta(z) := \sum\limits_{n=1}^\infty n^{-z}$
has all its zeros in the critical strip
$\cS := \{z \ | \ 0 < \re(z) < 1 \}$
lying on the critical line
$\cL := \{ z \ | \ \re(z) = \frac12 \}$.}
\end{quote}
Now, $\zeta(z)$ obeys a remarkable functional equation
$\zeta(z) = 2^z \pi^{z-1} \sin(\frac12 \pi z) \Gamma(1-z) \zeta(1-z)$,
which inspired Riemann to define the $\xi$-function, which will be central to our discussions:
\begin{equation}\label{xi}
\xi(z) := \frac12 z(z-1) \pi^{-\frac{z}{2}} \Gamma(\frac{z}{2}) \zeta(z) ~.
\end{equation}
There are many advantages to considering $\xi(z)$ instead of $\zeta(z)$, which we now summarize.

\subsection{The Riemann $\xi$-Function}
First, note that $\zeta(z)$ has trivial zeros at all negative even integers, which are conveniently cancelled by the corresponding simple poles of the $\Gamma$-function.
Hence, $\xi(z)$ has only non-trivial zeros, all located within the critical strip, and affords an elegant Weierstra\ss\ product expansion, known as the {\it Hadamard product}:
\begin{equation}\label{hadamard}
\xi(z) = \xi(0) \prod_{\rho \in \cS} (1 - \frac{z}{\rho}) ~,
\end{equation}
where $\xi(0) = \frac12$, and $\rho$ are the non-trivial zeros of $\zeta(z)$ in $\cS$ since none of the other factors in the definition of $\xi(z)$ vanishes therein.
It is an obvious but important fact that $\cS$ extends to both the upper and the lower half-planes and that all the $\rho$ occur in {\it conjugate pairs} above and below the real line.
We shall denote $\cS_+$ as the critical strip above the real line and correspondingly $\cL_+$ as the upper critical line.

Second, the functional equation becomes particularly simple:
\begin{equation}\label{fe}
\xi(z) = \xi(1-z) ~.
\end{equation}
Indeed, the symmetry about the critical line of $\re(z) = \frac12$ becomes manifest.
Hence, not only are the zeros symmetric about the real axis, they are also symmetric about the critical line.
The Riemann Hypothesis postulates that all the zeros in fact lie on $\cL$.

Third, consider the conformal mapping $z \mapsto \frac{z}{z-1}$, which maps $\cL_+$ to the boundary circle $|z|=1$ of the unit disk and the entirety of the left half of
$\cS_+$, {\em i.e.}, $\{z \ | \ 0 < \re(z) < \frac12 \}$,
to the interior of the open unit disk.
Consider, therefore, the function
\begin{equation}\label{phi-fe}
\phi(z) := \xi(\frac{z}{z-1}) = \xi(\frac{1}{1-z}) ~,
\end{equation}
where we have used \eqref{fe} in the second equality.
Since all zeros are symmetric around the critical line, to consider the zeros of $\xi$ within $\cS$ therefore amounts to considering the zeros of $\phi(z)$ within the open unit disk.
Indeed, the Riemann Hypothesis would require that there be no such zeros and indeed that all the critical zeros are on the boundary circle.
Since the number of zeros of an analytic function $f(z)$ in a region $R$ is equal to
$\int_{\partial R} {\dd}z\ \frac{f'(z)}{f(z)}$,
we have that
\begin{proposition}
The Riemann Hypothesis is equivalent to $\frac{\phi'(z)}{\phi(z)}$ being analytic ({\em i.e.}, holomorphic and without poles) within the open unit disk $|z| < 1$.
\label{phi}
\end{proposition}

Fourth, a beautiful exact result relates the number of zeros up to height $T$ within $\cS$, and the $\xi$-function:
\begin{equation}\label{nt}
N(T) := \# \{ \rho \in \cS ~, 0 < \im(\rho) < T | \ \xi(\rho) = 0 \} =
\frac{1}{\pi} \im \log \xi (\frac12 + i T) ~.
\end{equation}
Briefly, this can be seen~\cite{e} as follows.
We have that, since there are no poles in the critical strip,
$N(T) = \frac{1}{2\pi i} \int_{\partial R} {\dd} s\ \frac{\xi'(s)}{\xi(s)} = \frac{1}{2\pi} \im \int_{\partial R} {\dd}s\ (\log \xi(s))'$,
where $R$ is the rectangular region in the critical strip up to height $T$, strictly,
$R = \{z \ | \ -\epsilon \le \re(z) \le 1 + \epsilon, \ 0 \le \im(z) \le T \}$.
(Note that no zero passes through the line $\im(z)=T$.)
By the symmetry of \eqref{fe}, we have that
$N(T) = 2 \frac{1}{2\pi} \im \int_{C} {\dd}s\ (\log \xi(s))'$,
where $C$ is the right half segment of $\partial R$ from $\frac12$ to $\frac12 + i T$, and whence \eqref{nt} since $\log \xi(s)$ is real at $\frac12$.

Extensive work has been done in estimating \eqref{nt}.
We have, using the definition \eqref{xi} for the first equality, that
\begin{equation}\label{eq:asym}
N(T) = \frac{1}{\pi} \vartheta(T) + 1 + \frac{1}{\pi} \im \log \zeta(\frac12 + i T) \sim \frac{T}{2\pi} \log \frac{T}{2\pi} - \frac{T}{2\pi} - \frac78 + \cO(\log T) ~,
\end{equation}
where historically $\vartheta(T) = \im \log \Gamma(\frac{i}{2} T + \frac14) - \frac{T}{2} \log \pi$ is known as the average part and $\frac{1}{\pi} \im \log \zeta(\frac12 + i T) \sim \cO(\log T)$ is the fluctuating part around the essentially $T \log T$ growth of the former.
This is the inspiration behind classical (the average) and quantum mechanical (the fluctuation) interpretations of the critical zeros~\cite{hp, berry}.

We note that $N(T)$ is a real step function\footnote{
We use the standard notation that the step function is defined as
$\theta(x - a) = \left\{
\begin{array}{ll}
1, & x \ge a \\
0, & x < a
\end{array}
\right.$. Its derivative is the delta-function $\delta(x-a) = \theta'(x-a)$.
Importantly, the integral of $\delta(x-a)$ over any finite interval containing $a$ on the real axis is normalized to equal to unity.
}, increasing by unity each time a new critical zero is encountered:
\begin{equation}\label{NT}
N(T) = \sum_{\rho \in \cS_+} \theta(T - \im \rho) = \sum_{j = 1}^\infty \alpha_j\ \theta(T - \mu_j) ~.
\end{equation}
It is vital to explain the above in detail.
We can explicitly write the upper critical zeros as $\rho_{j,k} = \sigma_{j,k} + i \mu_j$, indexed by $j, k\in \IZ_{>0}$, where $\sigma_{j,k}\in (0,1)$ and $\mu_j\in \IR_{>0}$.
The zeros are ordered so that $\mu_{j+1} > \mu_j$.
Crucially, as we do not assume the Riemann Hypothesis, the real part of $\rho$ need not be $\frac12$.
Furthermore, we do not assume simplicity of the zeros, which is as yet also unknown~\cite{conrey2}.
If, for example, we have a double root, we explicitly count this twice.
By the functional identity, zeros of the $\xi$-function not on the critical line are paired within the upper critical strip:
if $\sigma + i \mu$ is a zero, then so is $(1-\sigma) + i\mu$.
The index $k=1,\ldots,\alpha_j$ enumerates the zeros with the same imaginary part.
We order the zeros so that $\sigma_{j,k+1} \ge \sigma_{j,k}$.
The $\alpha_j$ then counts the number of zeros with imaginary part $\mu_j$ including the multiplicities of the roots.
If, for example, there are a pair of simple roots with imaginary part $\mu_j$, then in an epsilon interval around $\mu_j$, the counting function $N(T)$ jumps by two.
Contrariwise, if $\rho\in \cL_+$ is the only simple root with imaginary part $\mu_j$, then $N(T)$ jumps by one.
We know that the number of roots with imaginary part in the interval $(0,T)$ is finite.
Indeed, the asymptotics of the expression are given in \eref{eq:asym}.
In summary, $N(T)$ is a strictly increasing step function as we move up in height $T$ regardless of the Riemann Hypothesis.

The first few numerical values for $\mu_{j = 1, 2, 3, 4, \ldots}$ are approximately $14.134725142$, $21.022039639$, $25.010857580$, $30.424876126$, $\ldots$.
It is known that the first $10^{13}$ zeros lie on the critical line~\cite{gourdon}.
The Riemann Hypothesis is, of course, the statement that $\sigma_j = \frac12$ for all zeros.

The rewriting of $N(T)$ lends itself to a wonderful interpretation:
$N(T)$ is a cumulative density function defined over the critical strip.
In other words, its derivative, $\widetilde{\rho}(T) := N'(T)$ is a density function of distributions.
That is, one could conceive of a physical system whose energy levels (eigenvalues of the Hamiltonian) are thus distributed;
this is along the school of thought of Hilbert--P\'olya~\cite{hp}.
Of course, the resulting eigenvalue density is highly non-smooth, but is, rather, a sum of delta-functions:
\begin{equation}\label{density}
\widetilde{\rho}(T) := N'(T) = \sum_{j = 1}^\infty \alpha_j\ \delta(T - \mu_j) ~.
\end{equation}

\subsection{Li's Criterion}
Another striking property of the $\xi$-function was noted by Li~\cite{li} not too long ago.
Let $\{k_n\}$, for positive integers $n$, be defined by
\begin{equation}
k_n := \frac{1}{(n-1)!} \frac{\dd^n}{\dd z^n} \left[ z^{n-1} \log \xi(z) \right]_{z=1} ~;
\end{equation}
then we have that
\begin{theorem}\label{li}
[Li's Criterion]
The Riemann Hypothesis is equivalent to the condition that $k_n \ge 0$ for all $n \in \IZ_{>0}$.
\end{theorem}

Li also showed two equivalent ways of writing these numbers, namely
\begin{eqnarray}
\label{kn-phi}
\frac{\phi'(z)}{\phi(z)} &=& \sum_{n=0}^\infty k_{n+1} z^n ~; \\
\label{kn-rho}
k_n &=& \sum_{\rho \in \cS} \left[ 1 - \left(1 - \frac{1}{\rho}\right)^n \right] ~,
\end{eqnarray}
where we recall that
\begin{equation}\label{phi-a}
\phi(z) := \xi(\frac{1}{1-z}) := 1 + \sum_{j=1}^\infty a_j z^j ~.
\end{equation}
Therefore, \eqref{kn-phi} means that $k_n$ are simply the Taylor expansion coefficients of $(\log \xi (\frac{1}{1-z}))'$.
We stress that in order for this Taylor expansion to make sense, the radius of convergence must be such that there are no poles within the enclosed disk (within the unit disk) thus defined. That is, we cannot allow a zero to occur for the denominator. Let a zero be $\rho := \sigma + i \ \mu$, then the first $z_0$ value for which $\frac{1}{1-z_0} = \rho$ will restrict the radius of convergence. Whence $|z_0| = \frac{|(\sigma-1)+i \mu|}{|\sigma + i \mu|}$.
In \cite{li}, to be safe, a worst bound is assumed, where $\sigma$ is taken to be 1 and $|z_0| = \frac14$.

On the other hand, \eqref{kn-rho} is also important~\cite{li, bl};
note that the sum is real because all critical zeros $\rho$ occur in conjugate pairs.
Furthermore, note that the modulus
$\left| 1 - \frac{1}{\rho} \right|$
can neither exceed nor be exceeded by unity since the summand
$1 - \left(1 - \frac{1}{\rho}\right)^n$
in the former case grows polynomially in $n$ and the sum diverges, while it tends to $1$ in the latter case and the sum also diverges.
Indeed, if the Riemann Hypothesis held, then for the $j$-th pair of conjugate critical zeros
$\left| 1 - \frac{1}{\rho} \right| = \left| 1 - \frac{1}{\frac12 \pm i \mu_j} \right|$,
which is exactly equal to $1$ and the sum in \eqref{kn-rho} converges.
Moreover, the unimodularity allows us to set
$1 - \frac{1}{\frac12 \pm i \mu_j} := \exp(\pm i \theta_j)$,
whence we would have
\begin{equation}\label{kn-li}
k_n = \sum_{j=1}^\infty (1 - \exp(i n \theta_j)) + (1 - \exp(-i n \theta_j)) = 2 \sum_{j=1}^\infty ( 1 - \cos(n \theta_j)) \ge 0 ~.
\end{equation}

Now, for the converse.
Li explicitly calculated the coefficients $a_j$ in \eqref{phi-a} to be
\begin{equation}
a_j = 4 \sum_{p=1}^j { j-1 \choose j-p } \frac{1}{p!}\ \int_p^\infty {\dd}x\ [x^{3/2} \psi'(x)]' \left(\frac12 \log x\right)^p
[1 + (-1)^p x^{-1/2}] \in \IR_{>0} ~,
\end{equation}
where $\psi(x) := \sum\limits_{n=1}^\infty e^{-\pi n^2 x}$.
Using \eqref{kn-phi} and \eqref{phi-a}, one has the recursion relation that
\begin{equation}
k_n = n\, a_n - \sum_{j=1}^{n-1} k_j\, a_{n-j} ~.
\end{equation}
If all $k_n \ge 0$, then the above recursion would imply that $k_n \le n\, a_n$ for all $n \in \IZ_{>0}$, whence for $z$ in the unit disk,
\begin{equation}
\left|\frac{\phi'(z)}{\phi(z)}\right| \le
\sum_{n=1}^\infty \left|k_n z^{n-1} \right| \le
\sum_{n=1}^\infty n\, a_n \left|z \right|^{n-1} =
\phi'(|z|) < \infty ~.
\end{equation}
That is, $\frac{\phi'(z)}{\phi(z)}$ is analytic in the open unit disk and by Proposition~\ref{phi}, Riemann Hypothesis holds.
Thus, the necessary and sufficiency together implies Theorem~\ref{li}.

\enlargethispage\baselineskip

Explicitly, one can Taylor expand to find the Li coefficients.
The first few are:
\begin{eqnarray}
\nn
k_1 &=& \frac12 \left( 2 + \gamma + \log\frac{1}{4\pi} \right) ~, \\
k_2 &=& 1 + \gamma - \gamma^2 + \frac{\pi^2}{8} + \log\frac{1}{4\pi} - 2\gamma_1 ~, \\
\nn
k_3 &=& 1 - 3\gamma^2 + \gamma^3 + \frac{3\pi^2}{8} + \frac32 \log\frac{1}{4\pi} - \frac{1}{16}\psi_2(1) - 6\gamma_1 + \gamma \left( \frac32 + 3\gamma_1 \right) + \frac32\gamma_2 - \zeta(3) ~,
\end{eqnarray}
where $\gamma$ is the Euler--Mascheroni constant, $\gamma_j$ are the Stieltjes constants\footnote{
We recall the definition of the Stieltjes constants.
Consider the series expansion of $\zeta(z)$ about $z=1$:
\begin{eqnarray}
\zeta(z) &=& \frac{1}{z-1} + \sum_{n=0}^\infty \frac{(-1)^n}{n!}\ \gamma_n\ (z-1)^n ~, \nn \\
\gamma_n &=& \lim_{m\to\infty} \left[ \sum_{k=1}^m \frac{(\log k)^n}{k} - \frac{(\log m)^{n+1}}{n+1} \right] ~. \nn
\end{eqnarray}
The $\gamma_n$ are then the Stieltjes constants.
We can as well define
\begin{equation}
\gamma_n = \frac{\dd^n}{\dd z^n}\zeta(z) - \left. \frac{(-1)^n\ n!}{(z-1)^{n+1}} \right|_{z=1} ~. \nn
\end{equation}
In particular, $\gamma_0 = \gamma$, the Euler--Mascheroni constant.},
and $\psi_n(z) = \frac{\dd^{n+1}}{\dd z^{n+1}} \log\Gamma(z)$ is the polygamma function. Indeed, we present some of the first numerical values of these coefficients in Table~\ref{t:li}.

\section{Density Function for the Distribution of Critical Zeros}\setall
Thus armed with all necessary ingredients, let us proceed to rewrite the Li coefficients in a suggestive form. We will do so {\it formally} in two ways and point out the subtleties involved at the end of the manipulations.

\subsection{A Logarithmic Expansion}
First, integrating \eqref{kn-phi} and using the definition as well as the functional equation \eqref{phi-fe}, we have that, for $z$ in the unit disk,
\begin{equation}\label{xi-expansion}
\log \xi(\frac{z}{z-1}) =
\log \xi(\frac{1}{1-z}) =
\sum_{n=1}^\infty k_{n} \int {\dd}z\ z^{n-1} =
-\log 2 + \sum_{n=1}^\infty \frac{k_{n}}{n} z^n ~,
\end{equation}
where $-\log 2$ is easily checked to be the constant of integration.

We note, however, that the above expansion has radius of convergence strictly less than unity, whereby excluding the unit circle, to which, crucially, $\cL$ is mapped under our conformal transformation $z \mapsto \frac{z}{z-1}$.
Therefore, it is imperative to analytically continue.
Let us start with \eqref{xi-expansion} and rewrite the expansion about the point $z=-1$. Note that the region of convergence for this is an open of circle of radius two centered about $z=-1$, which, in particular, encloses the entirety of the closed unit disk, except the point $z=1$, where there is a pole for $\phi(z) = \xi(\frac{1}{1-z})$.
\comment{
This is important since under our conformal map $z \mapsto \frac{z}{z-1}$, the critical line is mapped to the boundary unit circle centered at the original, on which \eqref{xi-expansion} fails to converge.}
Therefore, we have that
\begin{equation}\label{xi-exp2}
\log \xi(\frac{1}{1-z}) =
\log \frac12 + \sum_{n=1}^\infty \frac{k_{n}}{n} z^n =
b_0 + \sum_{n=1}^\infty b_n (z+1)^n ~,
\end{equation}
where we have written the two expansions together for comparison.
Indeed, as with the Li coefficients, all $b_n$ are clearly real.
Note that
\begin{equation}
b_0 := \log \frac{-\Gamma(\frac14)\zeta(\frac12)}{8\pi^{\frac14}} ~,
\quad
b_1 := \frac18 \left( -\log\pi + \psi_0(\frac14) + 2[\log\zeta(\frac12)]' \right) = 0 ~,
\end{equation}
as well as the curious identity $b_2 = b_3$.

Expanding $(z+1)^n$, we readily obtain an expression for the Li coefficients in terms of the new expansion coefficients $b_n$:
\begin{equation}\label{kn-an}
k_n = n \sum_{j=n}^\infty {j \choose n} b_j ~.
\end{equation}
\comment{
Therefore, showing that all $b_n \ge 0$ suffices to imply that all $k_n \ge 0$.
}
We now wish to solve for the $k_n$ explicitly and demonstrate positivity.

We first see, using the counting formula \eqref{nt} and the expansion \eqref{xi-exp2}, that
\begin{equation}\label{n-mu}
N(\mu) = \frac{1}{\pi} \im \log \xi(\frac12 + i \mu) =
\sum_{n=1}^\infty \frac{b_n}{\pi} \im \left(\frac{2\mu+i}{2\mu-i} + 1\right)^n ~,
\end{equation}
where we have used the substitution $\frac12 + i \mu = \frac{1}{1-z}$, or $z = \frac{2\mu+i}{2\mu-i}$.
Then, since
\begin{equation}
\im \left(\frac{2\mu+i}{2\mu-i} + 1\right)^n =
\frac{(4\mu)^n}{(4\mu^2+1)^{\frac{n}{2}}} \sin (n \tan^{-1} \frac{1}{2\mu}) =
2^n \cos^n\theta \ \sin(n \theta) ~, \qquad
\cos \theta := \frac{2\mu}{\sqrt{4\mu^2+1}} ~,
\end{equation}
we have that
\begin{equation}\label{N-mu}
\pi N(\mu) = \sum_{n=1}^\infty b_n 2^n \cos^n \theta \sin (n \theta) ~.
\end{equation}

Next, we apply the integral identity\footnote{
We can show this using the elementary identity that
$2^n \cos^n \theta \sin (n \theta) = \sum\limits_{p=0}^n {n \choose p} \sin(2 p \theta)$,
as well as the orthogonality condition that
$\int_{\frac{\pi}{2}}^0 {\dd}\theta\ \sin(2 p \theta) \sin(2 m \theta) = -\frac14 \pi \delta_{pm}$, $p,m \in \IZ$.
To our knowledge, \eref{eq:identity} first appeared in D.~Bierens de Hahn's
{\it Nouvelles tables d'int\'egrales d\'efinies}, Leiden: P.~Engels (1867),
published one year after Riemann's death.
}
\begin{equation}\label{eq:identity}
\int_0^{\frac{\pi}{2}} {\dd}\theta\ \cos^n\theta \sin (n\theta) \sin (2m\theta) = \frac{\pi}{2^{n+2}} {n \choose m} ~, \quad m,n \in \IZ_{>0} ~.
\end{equation}
Therefore, multiplying both sides of \eqref{N-mu} by $\sin (2m\theta)\ {\dd}\theta$ and integrating, we find that
\begin{equation}
\int_0^{\frac{\pi}{2}} {\dd}\theta\ \pi N(\mu)\ \sin(2 m \theta) = \sum_{n=1}^\infty b_n {n \choose m} \frac{\pi}{4} ~.
\end{equation}
Whence,
\begin{equation}
\sum_{n=1}^\infty b_n {n \choose m} =
4 \int_0^{\frac{\pi}{2}} {\dd}\theta\ N(\mu)\ U_{m-1}(\cos(2 \theta))\ \sin(2 \theta) ~,
\end{equation}
where we have used the definition of the Chebyshev polynomial of the second kind, that
\begin{equation}\label{chebyU}
U_{m-1}(\cos \phi) := \frac{\sin(m \phi)}{\sin \phi} ~.
\end{equation}
Finally, using that
$\cos(2 \theta) = \cos^2 \theta - \sin^2 \theta = \frac{4\mu^2-1}{4\mu^2+1}$,
that $\sin(2 \theta)\ {\dd}\theta = -2 \cos \theta\ {\dd}(\cos \theta)$, and that
as $\mu$ proceeds from $0$ to $\infty$, $\theta$ subtends an angle from $\frac{\pi}{2}$ to $0$, we find:
\begin{equation}
\sum_{n=1}^\infty b_n {n \choose m} =
8 \int_0^\infty {\dd}\mu\ N(\mu)\ U_{m-1}(\frac{4\mu^2-1}{4\mu^2+1}) \frac{2\mu}{\sqrt{4\mu^2+1}}\frac{2}{(4\mu^2+1)^{\frac32}} ~.
\end{equation}

However, recognizing the left hand side from \eqref{kn-an}, we arrive at the conclusion\footnote{
Curiously, if we did not worry about the region of convergence and proceeded {\it formally} with \eqref{xi-expansion}, we can obtain the same result.
Indeed, we can rewrite the function \eqref{nt} as, for $\mu \in \IR_{\ge 0}$,
\[
N(\mu) = \frac{1}{\pi} \im \log \xi(\frac12 + i \mu) =
\sum_{n=1}^\infty \frac{k_{n}}{\pi n} \im \left(\frac{2\mu+i}{2\mu-i} \right)^n ~,
\]
where we have set $\frac12 + i \mu = \frac{1}{1-z}$, or $z = \frac{2\mu+i}{2\mu-i}$ for the expansion and where we have used the fact that $k_n$ are all real.
Clearly, $\frac{2\mu+i}{2\mu-i}$ is unimodular and resides on the unit circle.
Using \eqref{chebyU}, we can simplify the expansion parameter:
\[
\im \left(\frac{2\mu+i}{2\mu-i} \right)^n =
\im e^{i \tan^{-1}(\frac{4\mu}{4\mu^2-1}) n} =
\sin \left( \tan^{-1}(\frac{4\mu}{4\mu^2-1}) n \right) =
\frac{4\mu}{4\mu^2+1} U_{n-1} ( \frac{4\mu^2-1}{4\mu^2+1} ) ~.
\]
Next, we recall that the Chebyshev polynomials prescribe an orthonormal basis over $[-1,1]$:
\[
\int_{-1}^1 {\dd}x\ U_n(x) U_m(x) \sqrt{1-x^2} = \frac{\pi}{2} \delta_{mn} ~,
\]
which, when transformed to $x := \frac{4\mu^2-1}{4\mu^2+1}$, becomes
\[
\int_0^\infty {\dd}\mu\
U_n ( \frac{4\mu^2-1}{4\mu^2+1} )
U_m ( \frac{4\mu^2-1}{4\mu^2+1} )
\frac{4\mu}{4\mu^2+1}
\frac{16\mu}{(4\mu^2+1)^2}
= \frac{\pi}{2} \delta_{mn} ~.
\]
Therefore, this allows for the inversion of the formula for $N(\mu)$ above by integration of both sides, and we arrive once more at \eqref{kn-int}.
}
that for all $m\in \IZ_{>0}$:
\begin{equation}\label{kn-int}
k_m = 32 m \int_0^\infty {\dd}\mu\ \frac{\mu}{(4\mu^2+1)^2}\ N(\mu)\ U_{m-1}(\frac{4\mu^2-1}{4\mu^2+1}) ~.
\end{equation}

This is beautiful and remarkable integral expression for the Li coefficients.
We numerically evaluate some of the first of these numbers and find them to be indeed close to the required values of Li.
This is reassuring, and the results are presented in Table~\ref{t:li}.
\comment{
Crucially, note that we have not invoked the Riemann Hypothesis in the above steps, but only the cumulative density function $N(\mu)$ for $\mu$ measuring the height within $\cS_+$ above the real axis.
}

Let us now proceed to simplify the integral \eqref{kn-int}.
First, recall that Chebyshev polynomials of the second kind are related to the Chebyshev polynomials of the first kind by the equation
\begin{equation}\label{int-TU}
\int {\dd}x \ U_n(x) = \frac{1}{n+1} T_{n+1}(x) ~.
\end{equation}
Next, integration by parts $\int_a^b {\dd}\mu\ f(\mu)\, g'(\mu) = [f(\mu)\, g(\mu)]_a^b - \int_a^b {\dd}\mu\ f'(\mu)\, g(\mu)$ suggestively compels us to define
\begin{equation}
f(\mu) := N(\mu) ~; \qquad
g'(\mu) := \frac{32m \mu}{(4\mu^2+1)^2} U_{m-1}(\frac{4\mu^2-1}{4\mu^2+1})
\Longrightarrow
g(\mu) = 2(T_m(\frac{4\mu^2-1}{4\mu^2+1}) - 1) ~.
\end{equation}
Note that we have added $-2$ as the integration constant for $g(\mu)$ for reasons shortly to be clarified.

The integral thus becomes\footnote{
One could also perform the integral without recourse to delta-functions and generalized analysis. The expression \eqref{kn-int} becomes, by \eqref{NT},
\begin{eqnarray}
\nn
k_m &=& 32 m \sum_{j=1}^\infty \alpha_j \int_0^\infty
{\dd}\mu\ \theta(\mu - \mu_j)
\frac{\mu}{(4\mu^2+1)^2} U_{m-1}(\frac{4\mu^2-1}{4\mu^2+1})
= 2 m \sum_{j=1}^\infty \alpha_j \int_{\mu_j}^\infty
{\dd}\mu\ \frac{16\mu}{(4\mu^2+1)^2} U_{m-1}(\frac{4\mu^2-1}{4\mu^2+1}) \\
&=& 2 m \sum_{j=1}^\infty \alpha_j \left[
\frac{1}{m} T_{m}(y)
\right]_{\frac{4\mu_j^2-1}{4\mu_j^2+1}}^1 =
2 \sum_{j=1}^\infty \alpha_j\ (1 - T_m(\frac{4\mu_j^2-1}{4\mu_j^2+1})) ~. \nn
\end{eqnarray}
}
\begin{equation}
k_m = \left. 2 N(\mu)(T_m(\frac{4\mu^2-1}{4\mu^2+1}) - 1) \right]_0^\infty -
2 \int_0^\infty {\dd}\mu\ \widetilde{\rho}(\mu) (T_m(\frac{4\mu^2-1}{4\mu^2+1}) - 1) ~,
\end{equation}
where, importantly, we have used \eqref{density} for the explicit form of
$\widetilde{\rho}(\mu)$ and the defining property of the delta-function.
Next, note that $N(0) = 0$ since the first critical zero does not occur until $\mu > 14$, and that for any fixed $m$,
\begin{equation}
\lim\limits_{\mu \rightarrow \infty} N(\mu)
(T_m(\frac{4\mu^2-1}{4\mu^2+1}) - 1) \sim
m^2(\frac{-2}{4\mu^2+1})\ \mu \log \mu = 0 ~.
\end{equation}
Thus, the boundary terms vanish, and we have that
\begin{equation}\label{kn-sum}
k_m = 2 \sum_{j=1}^\infty \alpha_j\ (1 - T_m(\frac{4\mu_j^2-1}{4\mu_j^2+1})) ~,
\quad m \in \IZ_{>0} ~.
\end{equation}

Recalling the definition of the Chebyshev polynomial of the first kind, that
$T_m(\cos \theta) := \cos (m \theta)$ and remembering that all $\alpha_j$, regardless of actual value, are strictly positive integers, each summand in \eqref{kn-sum} is non-negative.
In fact, because $\frac{4\mu_j^2-1}{4\mu_j^2+1} < 1$ strictly for any imaginary part as we know the first $\mu_j$ is at least $14$,
\begin{equation}
T_m(\frac{4\mu_j^2-1}{4\mu_j^2+1}) < 1 \quad
\mbox{ for any given } m \in \IZ_{>0}
\mbox{ and for all } j \in \IZ_{>0} ~,
\end{equation}
so each summand is actually strictly positive.
In conclusion, this would imply that all Li's coefficients $k_m > 0$.
Indeed, we have numerically evaluated the first few $k_m$ coefficients given in \eqref{kn-sum}, using the first $10^4$ critical zeros, and see that they are very close to the results obtained from \eqref{kn-phi} and \eqref{kn-int}, and moreover constitute an increasing sequence of strictly positive numbers.
In particular, the above discussions would force all critical zeros to lie entirely on $\cL$.
In this case, treating any multiple roots distinctly, we see that \eqref{kn-sum} reduces precisely to \eqref{kn-li}, which saturates the lower bound.

\comment{
Thus, if $(1 - T_m(\frac{4\mu_j^2-1}{4\mu_j^2+1}))$ is non-negative for all $\mu_j$, then a lower bound on $k_m$ is obtained by setting $\alpha_j = 1$:
\begin{equation}\label{kn-sum2}
k_m \ge 2 \sum_{j=1}^\infty (1 - T_m(\frac{4\mu_j^2-1}{4\mu_j^2+1})) ~,
\quad m \in \IZ_{>0} ~.
\end{equation}
Crucially, we have arrived at the conclusion that $k_m$ are real and non-negative {\it without} (unlike in the necessary direction of the proof in Li's Criterion) assuming the Riemann Hypothesis.
Therefore, our arguments seem indeed to compel all critical zeros of the Riemann zeta function to lie upon the critical line.
}

We seem to have arrived the positivity of the Li's coefficients without the assumption of the Riemann Hypothesis!
This, of course, is not quite true:
the criticality of the zeros of the $\xi$-function has in fact implicitly been invoked.\footnote{
We are grateful to R.~Heath-Brown and H.~Bui for pointing this out to us.}
Let us re-examine the expansion \eqref{xi-expansion}.
If a zero of $\xi$ were ever encountered, a branch cut must be carefully chosen because of the logarithm.
Now, in setting $\frac12 + i \mu = \frac{1}{1-z}$ in \eqref{n-mu}, we have used the fact that $\mu$ is real, whereby making $z$ on the boundary of the unit disk to which the critical line is mapped.
That is, $z = \frac{2\mu+i}{2\mu-i}$ is always unimodular for real $\mu$.
Therefore, only when {\it no} zeros at all lie in the interior of the unit disk, to which the region inside $\cS$ but not on $\cL$ is mapped, do we not need to carefully find the branch.
This, then, would require the Riemann Hypothesis.
Nevertheless, we have {\it formally} arrived the integral formula \eqref{kn-int} which does not seem to depend on the precise location of the zeros and the subsequent choice of branch cut.
We will discuss possible implications of this later.

\begin{table}[!t!b!h]
\begin{center}
\begin{tabular}{||c||c||c|c||c|c||}
\hline\hline
$n$ & $k_n$ by expansion & $k_n$ by integral & Difference & $k_n$ by sum & Difference \\ \hline\hline
$1$ & $0.0230957$ & $0.0235290$ & $-1.876\%$ & $0.0229610$ & $0.5832\%$ \\ \hline
$2$ & $0.0923457$ & $0.0940796$ & $-1.878\%$ & $0.0918069$ & $0.5835\%$ \\ \hline
$3$ & $0.207639$ & $0.211543$ & $-1.880\%$ & $0.206427$ & $0.5838\%$ \\ \hline
$4$ & $0.368790$ & $0.375740$ & $-1.884\%$ & $0.366635$ & $0.5844\%$ \\ \hline
$5$ & $0.575543$ & $0.586417$ & $-1.889\%$ & $0.572175$ & $0.5851\%$ \\ \hline
$6$ & $0.827566$ & $0.843252$ & $-1.895\%$ & $0.822717$ & $0.5860\%$ \\ \hline
$7$ & $1.12446$ & $1.14585$ & $-1.903\%$ & $1.11786$ & $0.5870\%$ \\ \hline
$8$ & $1.46576$ & $1.49376$ & $-1.911\%$ & $1.45713$ & $0.5881\%$ \\ \hline
$9$ & $1.85092$ & $1.88645$ & $-1.920\%$ & $1.84001$ & $0.5895\%$ \\ \hline
$10$ & $2.27934$ & $2.32332$ & $-1.930\%$ & $2.26587$ & $0.5910\%$ \\ \hline
$11$ & $2.75036$ & $2.80373$ & $-1.940\%$ & $2.73406$ & $0.5926\%$ \\ \hline
$12$ & $3.26326$ & $3.32695$ & $-1.952\%$ & $3.24386$ & $0.5944\%$ \\ \hline
$13$ & $3.81724$ & $3.89222$ & $-1.964\%$ & $3.79448$ & $0.5964\%$ \\ \hline
$14$ & $4.41148$ & $4.49869$ & $-1.977\%$ & $4.38508$ & $0.5985\%$ \\ \hline
$15$ & $5.04508$ & $5.14550$ & $-1.990\%$ & $5.01477$ & $0.6007\%$ \\ \hline
$16$ & $5.71711$ & $5.83170$ & $-2.004\%$ & $5.68263$ & $0.6032\%$ \\ \hline
$17$ & $6.42658$ & $6.55632$ & $-2.019\%$ & $6.38765$ & $0.6057\%$ \\ \hline
$18$ & $7.17248$ & $7.31832$ & $-2.033\%$ & $7.12884$ & $0.6085\%$ \\ \hline
$19$ & $7.95374$ & $8.11664$ & $-2.048\%$ & $7.90512$ & $0.6114\%$ \\ \hline
$20$ & $8.76928$ & $8.95012$ & $-2.062\%$ & $8.71540$ & $0.6144\%$ \\ \hline
\hline
\end{tabular}
\end{center}
{\caption {\sf The Li coefficients evaluated numerically.
Here, ``by expansion'' means the Taylor expansion in \eqref{kn-phi}, ``by integral'' means the integral identity in \eqref{kn-int}, with cut-off $10^8$ for the upper bound, and ''by sum'' means the summation over Chebyshev polynomials in \eqref{kn-sum}, with cut-off as the first $10^4$ critical zeros.
We have also juxtaposed the percentage errors with respect to the expansion values for comparative purposes.
}\label{t:li}}
\end{table}

\subsection{A Delta-Function Expansion}
Can we evade the problem of logarithmic branch cuts?
Let us return to the original definition of the Li's coefficients from \eqref{kn-phi}:
\begin{equation}
\label{eq:temp1}
\frac{\phi'(z)}{\phi(z)} =
\frac{1}{(1-z)^2} \frac{\xi'(\frac{1}{1-z})}{\xi(\frac{1}{1-z})} =
\sum_{n=0}^\infty k_{n+1} z^n =
\sum_{n=0}^\infty (n+1) b_{n+1} (z+1)^n ~.
\end{equation}
Thus defined, the relation between the $k$ and $b$ coefficients in \eqref{kn-an} still holds.

Now, let us, with the aid of some non-standard analysis, reinvestigate the counting function $N(\mu) = \frac{1}{\pi} \im \log \xi(\frac12+ i \ \mu) = \sum\limits_j \alpha_j\ \theta(\mu - \mu_j)$, where $\mu_j$ is the imaginary part of the $j$-th zero in $\cS_+$, regardless of the Riemann Hypothesis, and $\alpha_j$ counts potential multiplicities.
Upon taking the derivative, this gives us a sum over delta-functions, or,
\begin{equation}
\label{eq:temp2}
\sum_{j=1}^\infty \alpha_j\ \delta(\mu - \mu_j) =
\frac{1}{\pi} \re \left[ \frac{\xi'(\frac12 + i\, \mu)}{\xi(\frac12 + i\, \mu)} \right] ~,
\end{equation}
where we have exchanged the derivative with the $\im$. 
\comment{
Does the derivative of an $\im \log$-function tend, in the correct limit, to a delta-function?
}

Again, using the conformal transformation $\frac12 + i \mu = \frac{1}{1-z}$, or $z = \frac{2\mu+i}{2\mu-i}$, we have
\begin{equation}
\label{eq:temp3}
\re \left[ \frac{\xi'(\frac12 + i\, \mu)}{\xi(\frac12 + i\, \mu)} \right] =
\sum_{n=0}^\infty (n+1) b_{n+1} \re[(1-\frac{2\mu+i}{2\mu-i})^2 (\frac{2\mu+i}{2\mu-i}+1)^n] ~.
\end{equation}

\comment{Have we pulled a fast one?
In \eref{eq:temp1}, when we take the derivative, we examine the change in the function $\xi$ as its argument $z$ varies in an $\epsilon$ disk.
In the formula for $N(\mu)$, $\mu$ is real and the imaginary part of $\mu$ is by definition zero.
In \eref{eq:temp2}, when we take the derivative, we examine the change in the function $\xi$ as the argument $\mu$ varies in an $\epsilon$ interval.
In \eref{eq:temp3}, we set the two formul\ae, which are obtained from different variations, equal to each other.
Is this allowed?}

Now, we must substitute the following relation:
\begin{eqnarray}
\nn
\re[ (1-\frac{2\mu+i}{2\mu-i})^2 (\frac{2\mu+i}{2\mu-i}+1)^n ]
&=& \re[ \frac{4}{4\mu^2+1} e^{i \tan^{-1} \frac{4\mu}{4\mu^2-1}} \frac{(4\mu)^n}{(4\mu^2+1)^{\frac{n}{2}}} e^{i\, n \tan^{-1} \frac{1}{2\mu}} ] \\
&=& (2^n\, \cos^n \theta) (4\, \sin^2 \theta) \cos (n \theta + \psi) \\
&=& 2^{n+2}\, \cos^n\theta\ \sin^2\theta\ \cos((n+2) \theta - \pi) ~, \nn
\end{eqnarray}
where we have used, upon defining the angles,
\begin{equation}
\label{eq:tantheta}
\tan\theta = \frac{1}{2\mu} ~, \quad \tan\psi = \frac{4\mu}{4\mu^2-1} \Longrightarrow 2 \theta = \psi + \pi ~.
\end{equation}
Simplifying $\cos((n+2)\theta-\pi) = -\cos((n+2)\theta)$, we find that
\begin{equation}\label{delta-kn}
\sum_{j=1}^\infty \alpha_j\ \delta(\mu - \mu_j) =
- \frac{1}{\pi} \sum_{n=0}^\infty (n+1)\ b_{n+1}\ 2^{n+2}\, \cos^n\theta\ \sin^2\theta\ \cos((n+2) \theta) ~.
\end{equation}

For each positive integer $m$, we shall hit both sides of \eref{delta-kn} with $4\sin^2(m\theta(\mu))\ {\dd}\mu$ and integrate the variable $\mu$ on the half-line $[0,\infty)$.
Recall that $2\sin^2(m\theta) = (1-\cos(2m\theta))$.
Thus, the left hand side becomes
\bea
\int_0^\infty {\dd}\mu \sum_{j=1}^\infty \alpha_j\ \delta(\mu-\mu_j)\ 2(1-\cos(2m\theta(\mu))) &=&
2 \sum_{j=1}^\infty \alpha_j\ (1 - T_m(\cos(2\theta(\mu_j))) \nn \\
&=& 2 \sum_{j=1}^\infty \alpha_j\ (1 - T_m(\frac{4\mu_j^2-1}{4\mu_j^2+1})) ~,
\eea
where we have used the defining relation of the Chebyshev polynomial of the first kind.

Now, we do the same on the right hand side.
From \eref{eq:tantheta}, $\mu\in[0,\infty)$ implies that $\theta\in[\frac{\pi}{2},0)$.
It is easy to check that $\dd\mu = -\frac{1}{2\sin^2\theta} \dd\theta$.
We as well note the definite integral identities
\bea
\int_0^{\frac{\pi}{2}} \dd\theta\ \cos^n\theta\ \cos((n+2)\theta) &=& 0 ~, \quad n\in \IZ_{\ge 0} ~, \\
\int_0^{\frac{\pi}{2}} \dd\theta\ \cos^n\theta\ \cos((n+2)\theta)\ \cos(2m\theta) &=&
\left\{
\ba{cc}
\frac{\pi}{2^{n+2}} {n \choose {m-1}} ~, & n > m-2 ~, \\
0 ~, & n \le m-2 ~.
\ea\right.
\eea
Putting these pieces together, we find
\bea
&& -\frac{1}{\pi} \sum_{n=0}^\infty (n+1)\ b_{n+1}\ 2^{n+2} \int_0^{\frac{\pi}{2}} \dd\theta\ \cos^n\theta\ \cos((n+2)\theta)\ (1-\cos(2m\theta)) \nn \\
&& \qquad = \sum_{n=m-1}^\infty b_{n+1}\ (n+1) {n \choose {m-1}} = m \sum_{n=m}^\infty b_m\ {n \choose m} = k_m ~.
\eea

Hence, we recover \eref{kn-sum}:
\be
k_m = 2 \sum_{j=1}^\infty \alpha_j\ (1 - T_m(\frac{4\mu_j^2-1}{4\mu_j^2+1})) ~,
\quad m \in \IZ_{>0} ~.
\ee

Again, we seem to have arrived at an expression which implies Li's positivity without assuming Riemann Hypothesis. However, re-examining \eqref{eq:temp1} reveals a subtlety.
The radius of convergence is, as pointed out in \cite{li}, for $|z| < \frac14$.
However, we have equated $z = \frac{2\mu+i}{2\mu-i}$ in \eqref{eq:temp3}, because $\mu$ is real, and $|z|=1$, whereby again making the expansion and subsequent derivations only formal.
Of course, one could argue that the delta-function involved is a generalized function and should be neither confined to nor validated by convergence, since it is, after all, itself a divergent quantity.
Nevertheless, one needs to rigorously define and manipulate the divergences carefully.

\subsection{Remarks}

Here we wish to collect some comments regarding our main formulas for the Li coefficients:
\be
k_n = 32 n \int_0^\infty {\dd}\mu\ \frac{\mu}{(4\mu^2+1)^2}\ N(\mu)\ U_{n-1}(\frac{4\mu^2-1}{4\mu^2+1})
= 2 \sum_{j=1}^\infty \alpha_j\ (1 - T_n(\frac{4\mu_j^2-1}{4\mu_j^2+1})) ~.
\ee
These formulas have been derived by formal manipulations of divergent series and then have been checked numerically.
The crucial question is whether they are true in a rigorous mathematical sense.
In particular, there might exist a contour deformation that takes the integral formula to the convergent region and leaves it unchanged.
We leave that question as open, and list these formulas as hypothetical assertions to be proven rigorously.
Li also shows that the Riemann Hypothesis for the Dedekind zeta function
$\zeta_K(s)$ is as well equivalent to the non-negativity of a sequence
of real numbers~\cite{li}.
We therefore conjecture that similar formulas to the expressions for
$k_n$ apply for the coefficients $\kappa_n$ in the series expansion of
$(\log\xi_K(\frac{1}{1-z}))'$.

Another question is whether these formulas assume the Riemann Hypothesis in some subtle way.
The derivations we have presented indeed suffer from this affliction, but it may be that an alternate derivation of this expression sidesteps the issue.
We will offer a few further comments on this score.
The crucial observation is that both the counting function $N(\mu)$ and the argument of the Chebyshev polynomials that appear in the integral formula for the Li coefficients know directly about factor of $\frac12$ only from the functional equation for the Riemann $\xi$-function.
One might also question whether the unimodularity of the trigonometric representation of the Chebyshev polynomials is the signal of a secret assumption of the Riemann Hypothesis, but once again we note that this unimodularity relies on the argument $\frac12+i\mu$ of the counting function $N(\mu)$ determined by the functional equation for the $\xi$-function.
This tempts us to think that the final integral and summation formulas for $k_n$ only know about the functional equation for the Riemann function $\xi$.
But this remains to be rigorously proven.

Finally, we wish to offer a couple of comments about the possible geometric and topological meaning of the integral and the summation formulas for $k_n$.
Here we wish to compare the integral and the summation formulas for the Li coefficients to the remarkably deep structures uncovered in the profound proofs of the Riemann Hypothesis for the case of finite fields by Weil, Deligne and others~\cite{weilconj}.
(An insightful summary of Weil's work is presented in the recent book~\cite{connes}, especially in chapter~4.)
The work of Li is also related to the seminal work of Weil by Bombieri and Lagarias~\cite{bl}.
We wish to note that the integral and summation formulas for $k_n$ discussed in this paper capture, at least heuristically, some of the crucial aspects discussed in these papers and books.

In particular, the equality of the integral and the summation formula for the Li coefficients reminds one of Weil's explicit formula~\cite{weil}.
This explicit formula has an interpretation in terms of an index theorem with deep meaning in algebraic geometry.
The integral formula for the Li coefficients could be understood as an index formula of the Atiyah--Singer type, which might have a profound topological interpretation in the case of the distribution of the Riemann zeros as discussed in Connes and Marcolli~\cite{connes}.
For example, the counting function appearing in the integral formula for $k_n$ might be seen to correspond to some generalized Todd genus and the Chebyshev polynomial to some generalized Chern character.
The summation formula for the Li coefficients on the other hand can be recast as an alternative sum, which is very similar to the Lefschetz fixed point formula which figures prominently in Weil's seminal work and in more recent topological discussions of the Riemann Hypothesis, as once again reviewed in~\cite{connes}.
Finally, even the argument of the Chebyshev polynomials that appears in the integral and the summation formulas for $k_n$ is suggestive of the existence of some non-trivial vector bundle structure over the space of integers, where the factor $\frac{4\mu_j^2-1}{4\mu_j^2+1}$ reminds one of the value of the Higgs field vortex configuration which determines the self-dual connection on that vector bundle.
(See the Section 7.6 of Chapter 4 of Connes and Marcolli~\cite{connes} for comparison.)
The summation formula then appears as the formula for the action of the compact (topological) Abelian Higgs theory, as prominently discussed in the case of compact quantum electrodynamics, as prominently discussed in the case of compact quantum electrodynamics \cite{polyakov}.

We intend to return to these fascinating structures in future work.


\section{Physical Perspectives}\setall
As we have noted in the Introduction, attempts to use a physical argument to explain the Riemann Hypothesis date back to Hilbert and P\'olya~\cite{hp}.
Were one to find a matrix
\begin{equation}\label{H}
H = \frac12 \cdot \id + i\cdot \cT ~,
\end{equation}
whose eigenvalues are all the zeros of $\xi(z)$ and if in addition the matrix ${\cal T}$ were Hermitian, then all the zeros would lie on the critical line $\cL$.
Therefore, one would wish to find the appropriate operator $H$ and its corresponding eigenfunctions, or, in physical terms, the Hamiltonian and its wave functions.

Berry and Keating~\cite{berry} pioneered a method to interpret $N(T)$ as a density of states.
However, they focused primarily on the ``average part'' of $N(T)$, {\em viz.}, the part that exhibits $T\log T$ growth.
We seek to use Li's Criterion in concert with our approach to further this interpretation.
The full expression $N(T) = \frac{1}{\pi}\im \log \xi(\frac12 + i T)$ inspires us to regard the $\xi$-function as a wave function,
\be\label{wave}
\xi(z) = r(z) e^{\frac{i \pi}{\hbar} S(z)} ~,
\ee
where $S(z)$ is the standard quantum mechanical phase and $r(z)$ is the modulus.
Subsequently $S(z) = \frac{\hbar}{\pi} \im \log \xi(z)$ indeed becomes $N(T)$ for $z = \frac12 + i T$.
In the Hamilton--Jacobi formalism of classical mechanics and its standard quantization, $S(z)$ in \eqref{wave} is simply the classical action.
We know that $N(T)$ is an integer.
Therefore, the action $S$ must be quantized, {\em i.e.}, $S = \hbar\, n$ for $n \in \IZ$,
and $S$ is therefore integral in units of $\hbar$.
This requirement is, of course, the Bohr--Sommerfeld quantization condition.

Furthermore, using \eqref{n-mu} and \eqref{int-TU}, as well as \eqref{kn-sum}, we have that, for the classical action ({\em i.e.}, dropping the factor of $\hbar$),
\be
S(z) = \frac{1}{\pi} \sum_{n=0}^\infty k_n {\dd}y_n ~, \quad
y_n := T_n(\frac{4z^2-1}{4z^2+1}) ~, \quad
k_n = 2 \sum_{j=1}^\infty \alpha_j \left(1 - T_n(\frac{4\mu_j^2-1}{4\mu_j^2+1})\right) ~.
\ee
This way of writing an action is well known from Hamilton--Jacobi's action-angle theory.
It means we have a classical system of an infinite but countable set of particles indexed by $n$, whose generalized coordinates are $y_n$ and conjugate momenta are $k_n$, with an energy $z$.
Therefore, we propose that the Li coefficients should be regarded as generalized momenta and are the action variables.
These must be positive in order to have a stable classical system.
If they were negative, the system would be unstable.
If they were complex, the system would be metastable.
Hence, the Li Criterion implies that our dynamical system is classically stable.

What can one say about the quantum mechanical picture?
Let us investigate the secular equation of the operator in \eqref{H}.
It should by construction have roots corresponding to the zeros of $\xi(z)$, that is,
\begin{equation}\label{eq:wkb}
\xi(z) = \det (z \cdot \id - H) ~.
\end{equation}
To arrive at the appropriate matrix description, we start by considering the expansion \eqref{xi-expansion}, which, combined with \eqref{kn-sum}, gives
\begin{equation}
\log\xi(\frac{1}{1-s}) = -\log 2+\sum_{n=1}^\infty \sum_{j=1}^\infty \frac{2\alpha_j}{n} \left[ 1 - T_n( \frac{4\mu_j^2-1}{4\mu_j^2+1} ) \right] s^n ~.
\end{equation}
Put $\cos\theta_j := \frac{4\mu_j^2-1}{4\mu_j^2+1}$.
Using the definition of the Chebyshev polynomial and reversing the sums, we obtain
\bea
\log\xi(\frac{1}{1-s})
&=& -\log 2 + \sum_{j=1}^\infty 2 \alpha_j \sum_{n=1}^\infty \left( \frac{1}{n} - \frac{1}{n}\cos(n \theta_j) \right) s^n \\
\nn
&=& -\log 2+\sum_{j=1}^\infty \alpha_j \left[ -2\log(1-s) + \log(1-s \ e^{i \theta_j}) + \log(1-s\ e^{-i \theta_j}) \right] ~.
\eea
Substituting $z = \frac{1}{1-s}$ and using that
$e^{\pm i \theta_j} = 1 - (\frac12 \pm i \mu_j)^{-1}$,
we find
\be\label{eq:xi1}
\xi(z) = \frac{1}{2} \prod_{j=1}^\infty \left( \frac{(z-(\frac12 + i\mu_j))(z-(\frac12 - i\mu_j))}{(\frac12 + i\mu_j)(\frac12 - i\mu_j)} \right)^{\alpha_j} ~,
\ee
which is simply a refined version of the Hadamard product, a reassuring check.\footnote{
The product formula \eref{eq:xi1} can also be obtained by using the integral formula \eqref{kn-int} derived in section 3.
Thus,
\[
\log\xi(\frac{1}{1-s}) = -\log 2+ 32\sum_{n=1}^\infty \int_0^{\infty} {\dd}\mu\ \frac{\mu}{(4 \mu^2 +1)^2}\ N(\mu)\ U_{n-1}(\frac{4\mu^2 -1}{4\mu^2 +1})\ s^n ~.
\]
We exchange the order of the integral and the sum and repeat the steps from the previous computation to obtain \eref{eq:xi1}.}
Note that each factor in the product obeys the functional identity.

Equation~\eref{eq:xi1} is to be compared with \eref{eq:wkb}.
We see that there exists a matrix such that its eigenvalues coincide with $\mu_j$, which is consistent with the Hilbert--P\'olya picture.
As the expression vanishes for zeros on the critical line, the refined Hadamard product is already in the form of the determinant of an infinite matrix, and the eigenvalues of $\cT$ are real.
Note the appearance of the $\frac12$ prefactor in \eref{eq:xi1}.
In a quantum mechanical system, it would be tempting to interpret this as the vacuum energy term, but we will give a different interpretation in what follows.

Knowing the energy levels (eigenvalues) of the Hamiltonian $H$ does not allow us to pin down what the latter actually is;
the previous discussion simply shows that it must exist.
We do know, however, that the partition function is
\be
Z(\beta) = {\rm tr}\left( e^{-\beta H} \right) =
\sum_{j=1}^\infty e^{-\beta\mu_j} ~,
\ee
for the inverse temperature $\beta$.
This object would define the statistical mechanics of Riemann zeros.

What can we say about the actual Hamiltonian?
Our discussion here is necessarily heuristic.
The physical interpretation of the Li formula suggests a system in which the momenta ($p$) and coordinates ($q$) are linearly related.
In the simplest case, this would, assuming factorization, correspond to the Hamiltonian for an inverted harmonic oscillator $H = \frac{1}{2} (p^2 - q^2)$, which, for example, features prominently in the non-perturbative definition of the two-dimensional critical string theory in terms of Matrix models.
Upon a further canonical transformation such a Hamiltonian would amount to $H = QP$.

Berry and Keating~\cite{berry}, Connes~\cite{connes}, and Sierra and Townsend~\cite{sierra} among others, have looked at precisely such a Hermitian quantum mechanical Hamiltonian.
For convenience, let us work in units where $\hbar= 1$.
We have
\be
H = \frac{1}{2} (QP+PQ) ~, \qquad P := - i \frac{d}{dQ} ~.
\ee
The $Q$ eigenfunctions of this scaling operator in the $Q, P$ phase space are:
\be
H \ \psi_E(q) = E \ \psi_E(q) ~, \qquad
\psi_E (q) = \frac{A}{q^{\frac{1}{2} - iE}}
\ee
with some appropriate normalization constant $A$.
The momentum state wave function is the Fourier transform:
\be\label{psip}
\Psi_E(p) = \int_{-\infty}^\infty {\dd}q\ e^{-i p\cdot q} \psi_E(q) =
\frac{A}{|p|^{\frac12 + i E}} 2^{i E} \frac{\Gamma(\frac14 + \frac12 i E)}{\Gamma(\frac14 - \frac12 i E)} ~.
\ee

A few further comments are in order.
\bi
\item First note that the $\frac12$ appears from the symmetrization $\frac12(QP+PQ)$ in the Hamiltonian.
It is the {\em same} $\frac12$ that appears in formula~\eref{nt} for the number of zeros in the upper critical strip with imaginary part less than $T$ and in \eref{eq:xi1}.
We see that this factor is different from the vacuum energy, even though it does conspire with the canonical commutation relations (the reason for the vacuum energy) to produce the $\frac12 +iE$ combination.

\item Classically $QP$ is the simplest Hamiltonian that yields the generator of the Mellin transform, {\em i.e.}, a power, or scaling function, and $\frac12-iE$ appears as the scaling dimension.
That is to say, the $\frac12(QP+PQ)$ Hamiltonian is selected as the required quantum scaling operator with eigenfunctions given by generators of the Mellin transform.

\item The importance of the Mellin transform is indicated in the deep study of Bombieri and Lagarias~\cite{bl} regarding Li's positivity condition.
In particular Bombieri and Lagarias show that Li's Criterion is simply the consequence of the famous Weil Criterion~\cite{weil}, which explicitly relies on properties of the inverse Mellin transform,
namely the fact that the convolution of the inverse Mellin transform of a function and its complement
(that is to say, its ``dual'' under the $s \mapsto 1-s$ map)
is equal to the sum of the inverse Mellin transforms.
The $QP$ Hamiltonian (even classically) enjoys this feature, provided that we identify in the formula for the Li coefficients $Q_n = e^{i n \theta_j}$ and $P_n = e^{-i n \theta_j}$.
The Hermitian $\frac12(QP+PQ)$ Hamiltonian is the quantum counterpart of the classical scaling Hamiltonian, which leads to the power law eigenfunctions that generate the Mellin transform.
\ei

We observe that there is a beautiful identity, valid for all $\re(z) > 0$, and in particular, within the critical strip, that $\zeta(z) \Gamma(z) (1- 2^{1-z}) = \int_0^{\infty} {\dd}x\ \frac{x^{z-1}}{e^{x}+1}$. Together with the duplication formula for the $\Gamma$-function, $\sqrt{\pi} \Gamma(z) = 2^{z-1} \Gamma(\frac{z}{2}) \Gamma(\frac{z}{2} + \frac{1}{2})$, one can write
\be\label{zetagamma}
\xi(z) =
f(z) \int_0^{\infty} {\dd}x\ \frac{x^{z-1}}{e^{x}+1} =
f(z) \sum_{n=0}^\infty (-1)^n \int_0^{\infty} {\dd}x\ e^{n x} x^{z-1} \ ,
\quad
f(z) := \frac{z(z-1)}{\pi^{\frac{1-z}{2}}(2^z-2)\Gamma(\frac{z+1}{2})} \ .
\ee
This tells us that the $\xi$-function, up to a factor which does not vanish in the critical strip, is the Mellin transform of a Fermi--Dirac distribution.
Note that, in the second equality we have trivially inserted the infinite geometric sum of $(-1)^n e^{n x}$ in order to compare with \eqref{psip}.

Thus, even though the momentum space wave functions of the $\frac12(QP+PQ)$ scaling Hamiltonian contain the $\Gamma$-function, we can, upon a sequence of rescaling transformations of the Hamiltonian, $QP \mapsto n\ QP$, obtain a formula involving the $\xi$-function.
The fact that this infinite alternate sum of transformed Hamiltonians is proportional to the $\xi$-function could be viewed as a confirmation of what is expected from \eref{eq:wkb} and \eref{eq:xi1}.
The characteristic polynomial may then be regarded as a WKB-like wave function of the appropriate Hermitian operator.

In other words, by comparing \eqref{zetagamma} to \eqref{psip}, we would like to interpret the left hand side as a particular limit of the characteristic polynomial $\det(z \cdot \id - H)$.
In this expression, the Hamiltonian is $H = \frac12(QP+PQ)$, with power-like wave functions $\psi_E(q) = A\ q^{-1/2 +iE}$, exact real eigenvalues $E$, and scaling dimensions $\frac12 - i E$.
This particular limit of the characteristic polynomial of $H$ is given as the rescaling $QP \mapsto n\ QP$, for an integer $n$, where we introduce the alternate summation over $n$.
The claim we make is that in this particular limit the characteristic polynomial of $H=\frac12(QP+PQ)$ has zeros at the critical zeros of $\xi(z)$.
This would be consistent with what we have written in \eref{eq:wkb}.

Connes~\cite{connes} has examined the action of the operator $e^{i q\cdot p}$, which can be viewed as the generator of classical symplectic transformations, albeit on an abstract non-commutative space.
In some sense, by mapping $QP \mapsto n\ QP$ and resumming alternately, we have ``fermionized'' the generator of the classical symplectic transformations.
Note also that $\sum\limits_{n=0}^\infty (-1)^n e^{-n x}$ can be viewed as a twisted partition function.
As such it can be understood in terms of determinants, as familiar from the physics reinterpretation of index formulas.
It is noteworthy that the alternate sums of traces feature prominently in Connes's work as well.
It would be interesting to understand this relation more precisely.

This peculiar limit of the characteristic polynomial of a Hermitian operator can be understood in direct analogy with what happens in the semiclassical WKB $\hbar \to 0$ limit of a simple harmonic oscillator.
In this case, the exact wave functions (the Hermite polynomials) reduce to the Airy function, which can be understood as a WKB limit of the characteristic polynomial of the Hamiltonian for the simple harmonic oscillator.
The equidistant spectrum of the harmonic oscillator gets rearranged to give the zeros of the Airy function in the $\hbar \to 0$ limit.
Note that the zeros of the Airy function have to stay collinear given their origin in the WKB limit of a Hermitian operator.
Similarly the continuum spectrum of the Hermitian $\frac12(QP+PQ)$ Hamiltonian gets rearranged into the collinear spectrum of the zeta function in the limit described in the preceding paragraph and in \eref{zetagamma}, which is different in detail, but similar in spirit to the canonical WKB limit.

Finally, we note that the continuous scaling dimension associated with the power law function that generates the Mellin transform should be preserved in this limit.
This could be responsible for the chaotic dynamical properties of critical Riemann zeros, as observed in manifold numerical studies.
The Fermi--Dirac distribution in \eref{zetagamma} might be the evidence for the repulsive (``fermionic'') nature of the Riemann zeros, which would fit the matrix model description of their correlation properties.

\newpage

\section*{Acknowledgments}
We are indebted to Jan de Boer, Hung Bui, Alexander Elgart, Roger Heath-Brown, Robert Karp, Rob Leigh, Jack Ng, Eric Sharpe, Tatsu Takeuchi, Chia Tze, and Alexandr Yelnikov for interesting conversations and questions regarding the material presented in this paper.
{\it Scientiae et Technologiae Concilio Y.-H.~H.\ hoc opusculum dedicat cum gratia ob honorem officiumque Socii Progressi collatum. Et Ricardo Fitzjames, Episcopo Londiniensis, ceterisque omnibus benefactoribus Collegii Mertonensis Oxoniensis quorum beneficiis pie, studiose, iucunde vivere licet, pro amore Catharinae Sanctae Alexandriae et ad Maiorem Dei Gloriam.}
\comment{Et pro rosa Hiberniae, fantasia Iberiae, delectatio Anatoliae, et oblectatio Caucasiae.
}
DM is supported in part by the U.S.\ Department of Energy under contract DE-FG05-92ER40677.
VJ is grateful to the LPTHE, Jussieu for their generous hospitality.
VJ and DM also wish to thank the magnanimity of the Warden, Fellows, and Scholars of Merton College, Oxford University.


\end{document}